\documentclass[a4paper]{spie}

\usepackage[sc, hang]{caption}
\usepackage{graphicx}
\usepackage{graphics}
\usepackage{makeidx}
\usepackage[pass]{geometry}
\usepackage{booktabs}
\usepackage{marvosym}
\usepackage[hang]{subfigure}
\usepackage{SIunits}
\usepackage{rotating}
\usepackage{csquotes}
\usepackage{url}
\usepackage{multicol}
\usepackage{multirow}
\usepackage{overpic}
\usepackage{amsmath}
\usepackage{commath}
\usepackage{bm}
\usepackage{rotating}
\usepackage{psfrag}
\usepackage{parallel}
\usepackage{fancyhdr}
\usepackage[breaklinks]{hyperref}

\usepackage[english]{babel} 
\usepackage{eso-pic,graphicx}
\usepackage{enumerate}
\usepackage{psfrag}
\usepackage{booktabs}
\usepackage{hyperref}      
\hypersetup{               
   pdftitle={mavisSPIE2022},    
   pdfauthor={GuidoAgapito},  
   colorlinks=true,      
   linkcolor=black,      
   anchorcolor=black,    
   citecolor=black,      
   filecolor=black,      
   menucolor=black,      
   pagecolor=black,      
   urlcolor=blue,         
   linktocpage=true
}

\usepackage{color, colortbl}
\definecolor{LightGray}{gray}{0.9}
\usepackage{diagbox}
\usepackage{array}
\newcolumntype{C}[1]{>{\centering\let\newline\\\arraybackslash\hspace{0pt}}m{#1}}

\title{MAORY/MORFEO and LIFT: can the low order wavefront sensors become phasing sensors?}

\author[a,c]{Guido Agapito}
\author[a,c]{Lorenzo Busoni}
\author[a,c]{Giulia Carlà}
\author[a,c]{Cédric Plantet}
\author[a,c]{Simone Esposito}
\author[b]{Paolo Ciliegi}
\affil[a]{INAF -- Osservatorio Astrofisico di Arcetri, Largo E. Fermi 5, 50125, Firenze, Italy}
\affil[b]{INAF -- Osservatorio di Astrofisica e Scienza dello Spazio di Bologna (OAS), via Gobetti 93/3, Bologna, Italy}
\affil[c]{ADaptive Optics National laboratory in Italy (ADONI)}

\authorinfo{Further author information:\\Guido Agapito, \Letter
 \hspace{0.5ex} guido.agapito@inaf.it\\}

\begin{document} 
\maketitle

\begin{abstract}
The Multiconjugate adaptive Optic Relay For ELT Observations (MORFEO, formerly known as MAORY) is the adaptive optics (AO) module for the Extremely Large Telescope (ELT) aimed at providing a 1 arcmin corrected field to the Multi-AO Imaging CamerA for Deep Observations (MICADO) and to a future client instrument. It should provide resolution close to the diffraction limit on a large portion of the sky and in a wide range of atmospheric conditions. Its ability to provide a flat wavefront must face the known aspect of the atmospheric turbulence and telescope environment, but also the final characteristic of a telescope still to be fully developed and built. In this work we focused on issues related to the segmentation of the telescope pupil (like low wind effect, residual phasing error at handover and control related issues), that could limit the system performance. MORFEO currently does not foresee a dedicated sensor to measure the phase step between adjacent mirror segments: in this work we study the possibility to use the low order wavefront sensors designed to sense and correct tip-tilt and focus as phasing sensors thanks to the linearized focal-plane technique (LIFT). 
\end{abstract}

\keywords{wavefront sensor, focal plane sensor, phasing sensor, segmented pupil, extremely large telescope, adaptive optics, multi conjugated adaptive optics, simulations}


\section{Introduction}
\label{sect:intro}  

Multiconjugate adaptive Optic Relay For ELT Observations\cite{2021Msngr.182...13C} (MORFEO, formerly know as MAORY) has many requirements to meet with the final goal of providing to 
The Multi-adaptive optics Imaging CamerA for Deep Observations \cite{2021Msngr.182...17D} (MICADO) resolution close to the diffraction limit on a large portion of the sky and in a wide range of atmospheric conditions.
One critical point that it will face is that MORFEO will work with ESO ELT\cite{2020SPIE11445E..1ET}, a telescope still to be fully developed and built.
In particular we do not know accurately which kind of mirror segmentation related issues there will be present.

In this work we focus on the possibility to add a petalometer to MORFEO to deal with differential ELT M4\cite{2019Msngr.178....3V} segment pistons.
The paper starts with a brief description of the segmentation related issue expected in MORFEO (Sec.~\ref{sec:segm}), then, in Sec.~\ref{sec:lift}, we present a focal plane sensing technique and how we used it to try to understand if the Low Orders WFSs\cite{2018SPIE10703E..4DB,Bonaglia2022SPIE} of MORFEO can be used to measure and correct for petal errors (see Sec.~\ref{sec:sim}).
Note that similar approaches are investigated also for the LTAO mode of HARMONI\cite{Bardou2022} and GMT\cite{Demers2022}.
Finally we discuss the results and the options we have (see Sec.~\ref{sec:consid}) and we evaluate in simulations one of these options (see Sec.~\ref{sec:scao_ass}).

\section{Segment phasing errors}\label{sec:segm}

ELT M4 segment phasing errors, known as petaling, could be present at the handover between the telescope and MORFEO or it could originate by low wind effect\cite{2015aoel.confE...9S,2018SPIE10703E..2AM} or control issues.
Actually we do not expect control issues, because MORFEO reconstruction and control is optimized to work with Von Karman turbulence and it should not develop petaling errors.

\begin{figure}[h!]
    \centering
    \subfigure[Aberration.\label{fig:lwe_aber}]
    {\includegraphics[width=0.34\columnwidth]{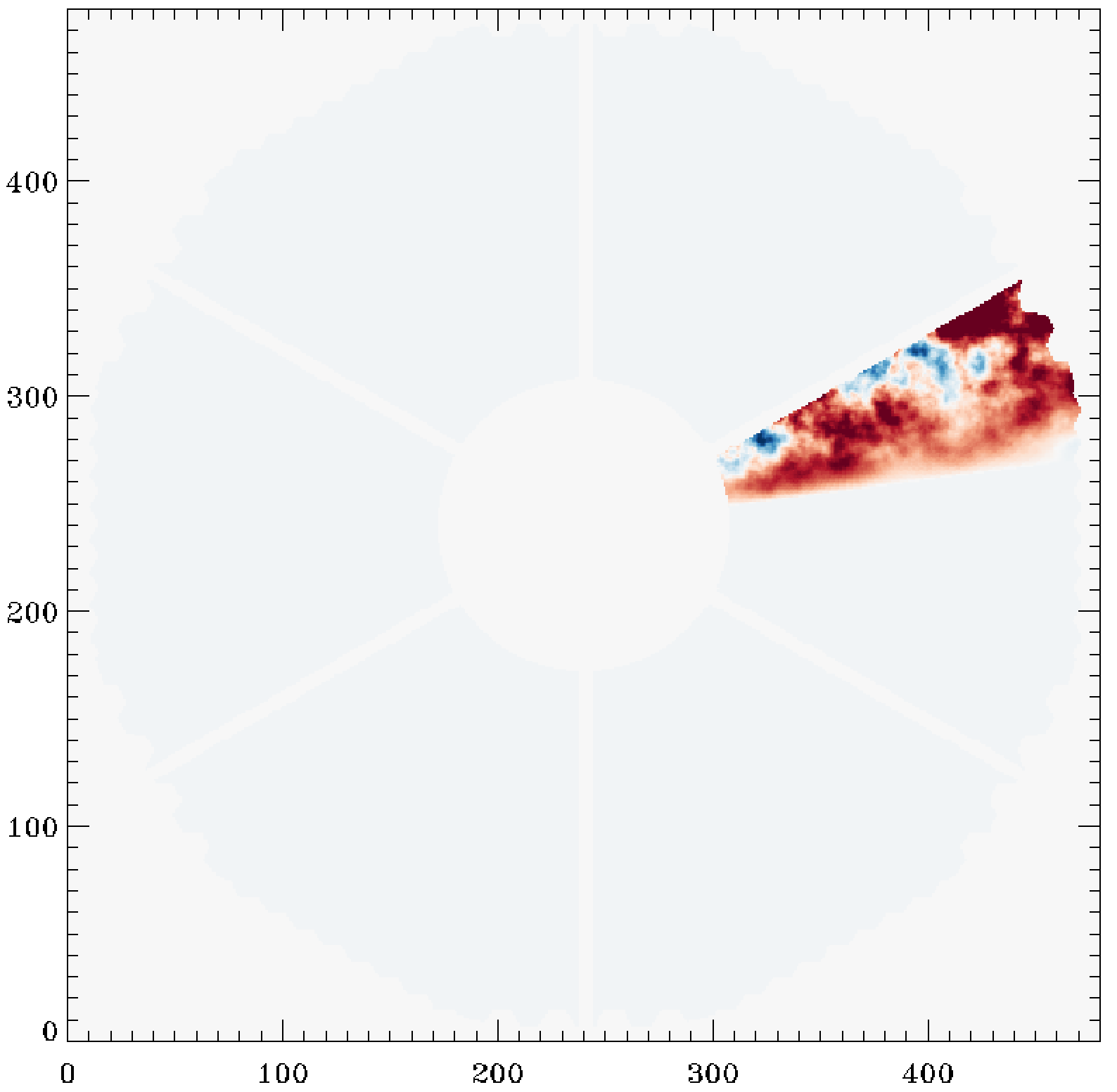}}
    \subfigure[M4 commands.\label{fig:lwe_comm}]
    {\includegraphics[width=0.47\columnwidth]{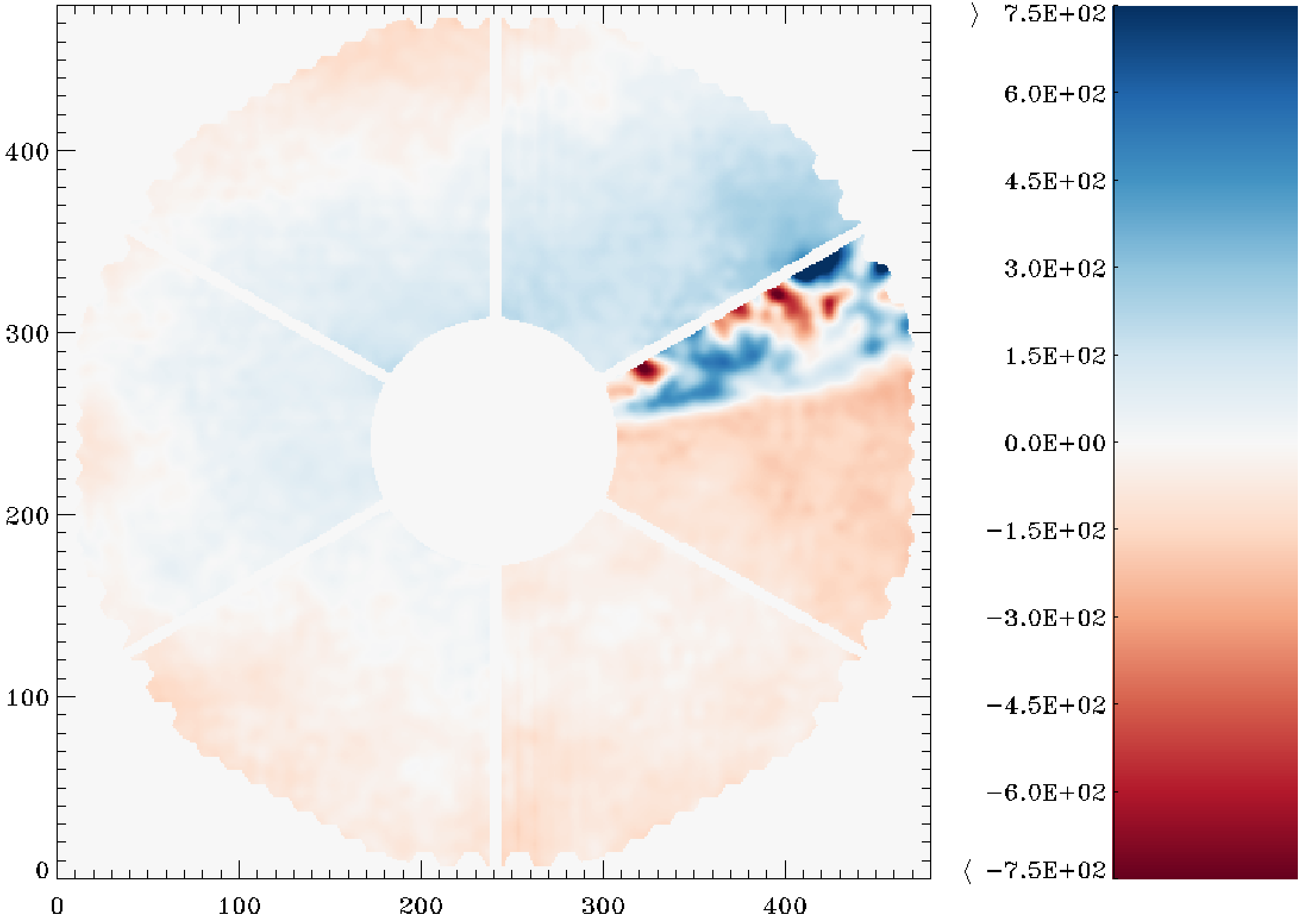}}
    \subfigure[Residual.\label{fig:lwe_res}]
    {\includegraphics[width=0.34\columnwidth]{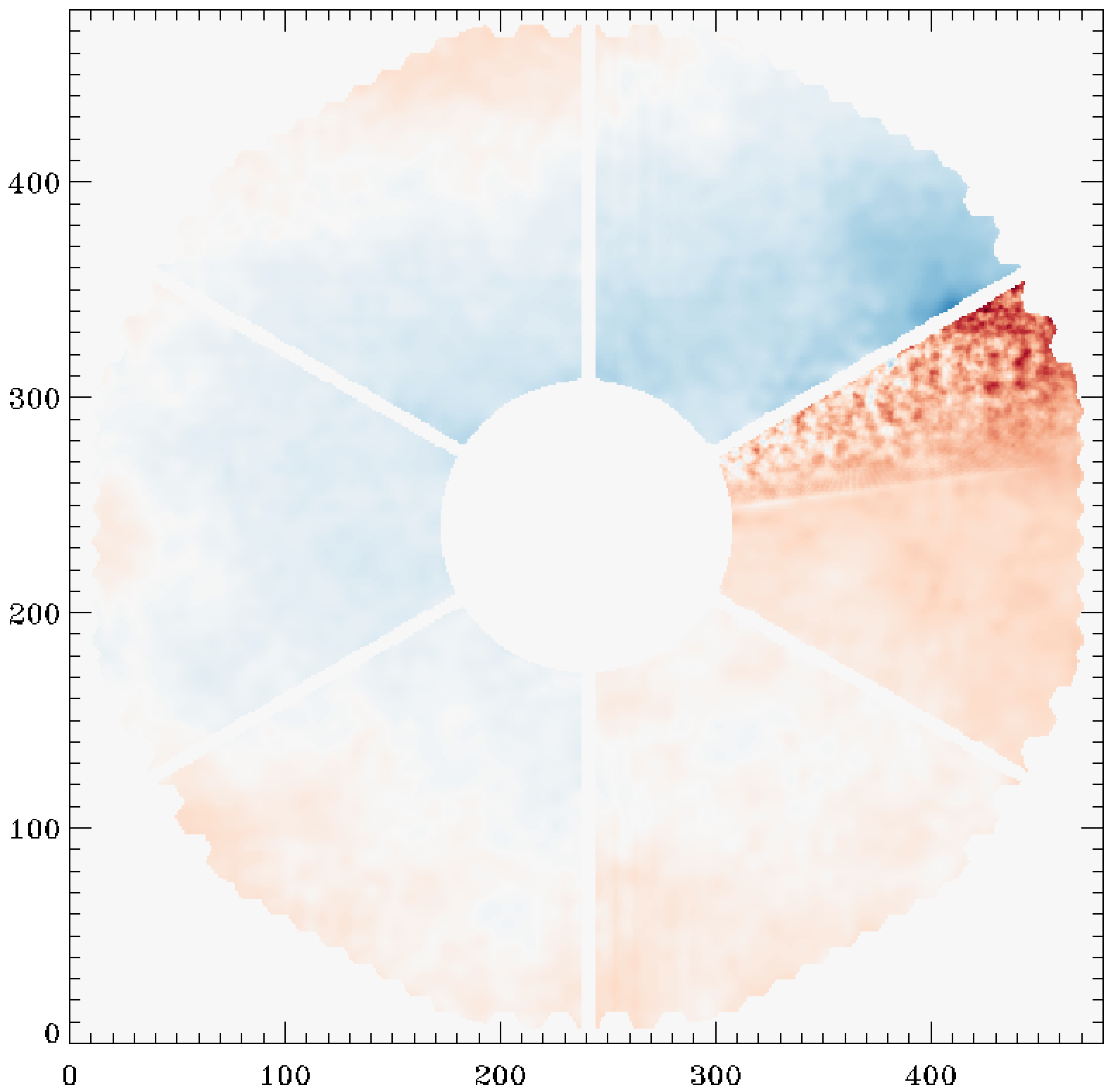}}
    \subfigure[Azimuthal plots.\label{fig:lwe_azim}]
    {\includegraphics[width=0.47\columnwidth]{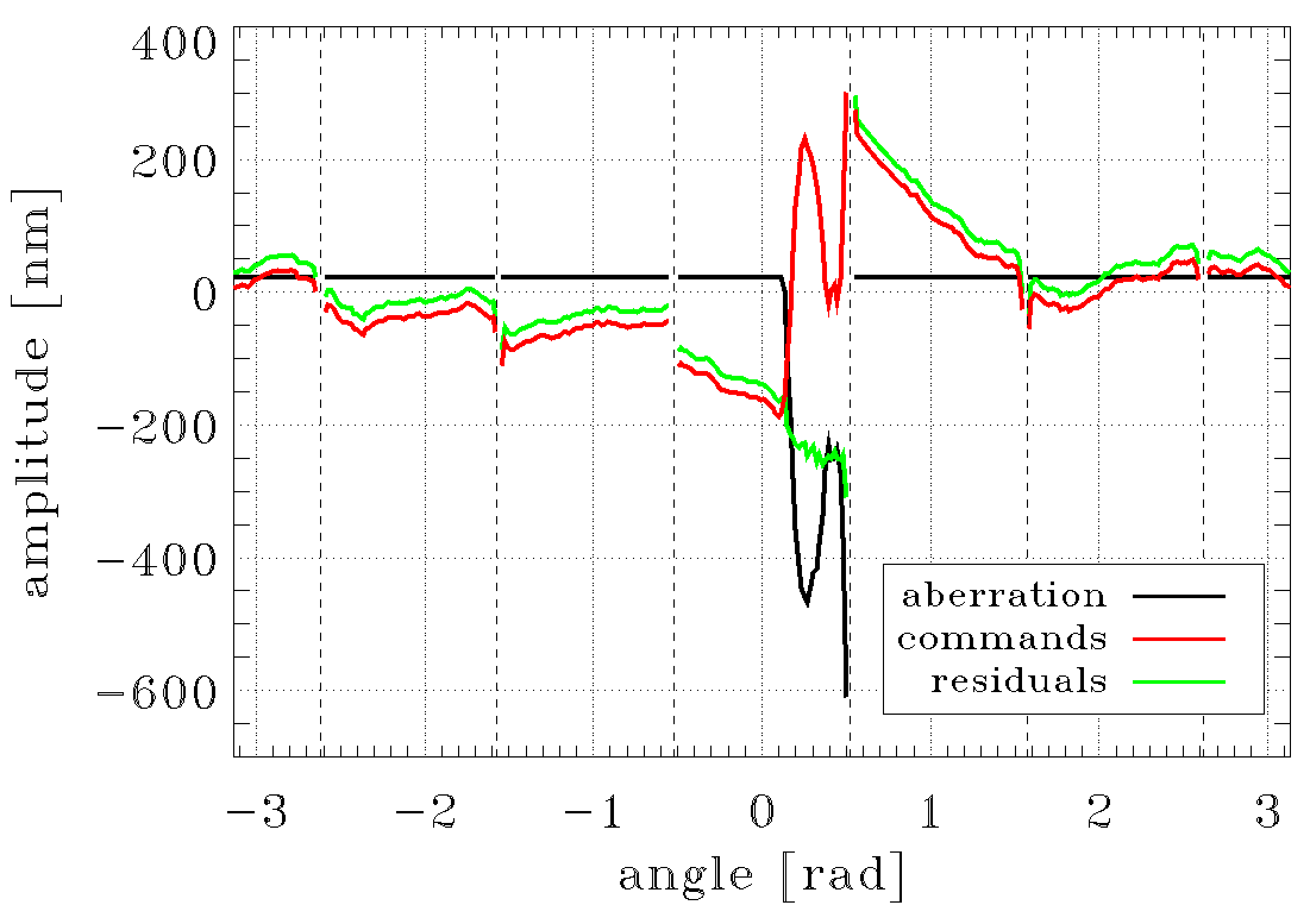}}
    \caption{Low Wind Effect simulation.\label{fig:lwe}}
\end{figure}

We decided to focus on low wind effect and we tested what happens with it in numerical simulations done with PASSATA\cite{doi:10.1117/12.2233963}.
We modeled it as a turbulence located on one side of a spider arm.
Please consider that it is an approximation and we consider that it is present on a single spider arm just to have a toy case to understand how the control of MORFEO deals with it.
We closed the loop on this turbulence and we saw that the residual phase is characterized by a petaling error.
The commanded shape of M4 is continuous, but combined with the low wind effect like turbulence we have a large jump between two petals (see Fig. \ref{fig:lwe}).
The WFSs of MORFEO are not sensitive to this error so it will not correct this aberration and this will impact on the science observations.

In this work we do not deepen the other possible phasing error sources, because we are just interested in showing that there is at least one source that requires a dedicated petal sensor.

\section{LIFT}\label{sec:lift}

We evaluated different options and we decided to focus on the linearized focal-plane technique \cite{Meimon:10} (LIFT) because it could be applied to the existing NGS 2$\times$2 SH WFSs.
One advantage of this technique in this case is that no reference aberration is required because pupil is not symmetric (pupil for each sub-aperture is a quarter of the ELT one) so we do not have the phase indetermination problem for even modes.

Please note that other focal plane sensing techniques have been studied for this kind of applications\cite{2017JATIS...3c9001L,2020SPIE11448E..6DV,2020A&A...639A..52B} and also other options are possible\cite{2006SPIE.6267E..32M,2008SPIE.7012E..3DP,2014SPIE.9148E..2MP,vanDam:16,2017A&A...603A..23J,2022JATIS...8b1513H,2022JATIS...8b1515H,2022arXiv220600295L}, but we would like to avoid substituting one WFS because this will impact directly the performance and the sky coverage of MORFEO. 

\section{Simulations}\label{sec:sim}

\begin{figure}[h]
    \centering
    \includegraphics[width=0.45\linewidth]{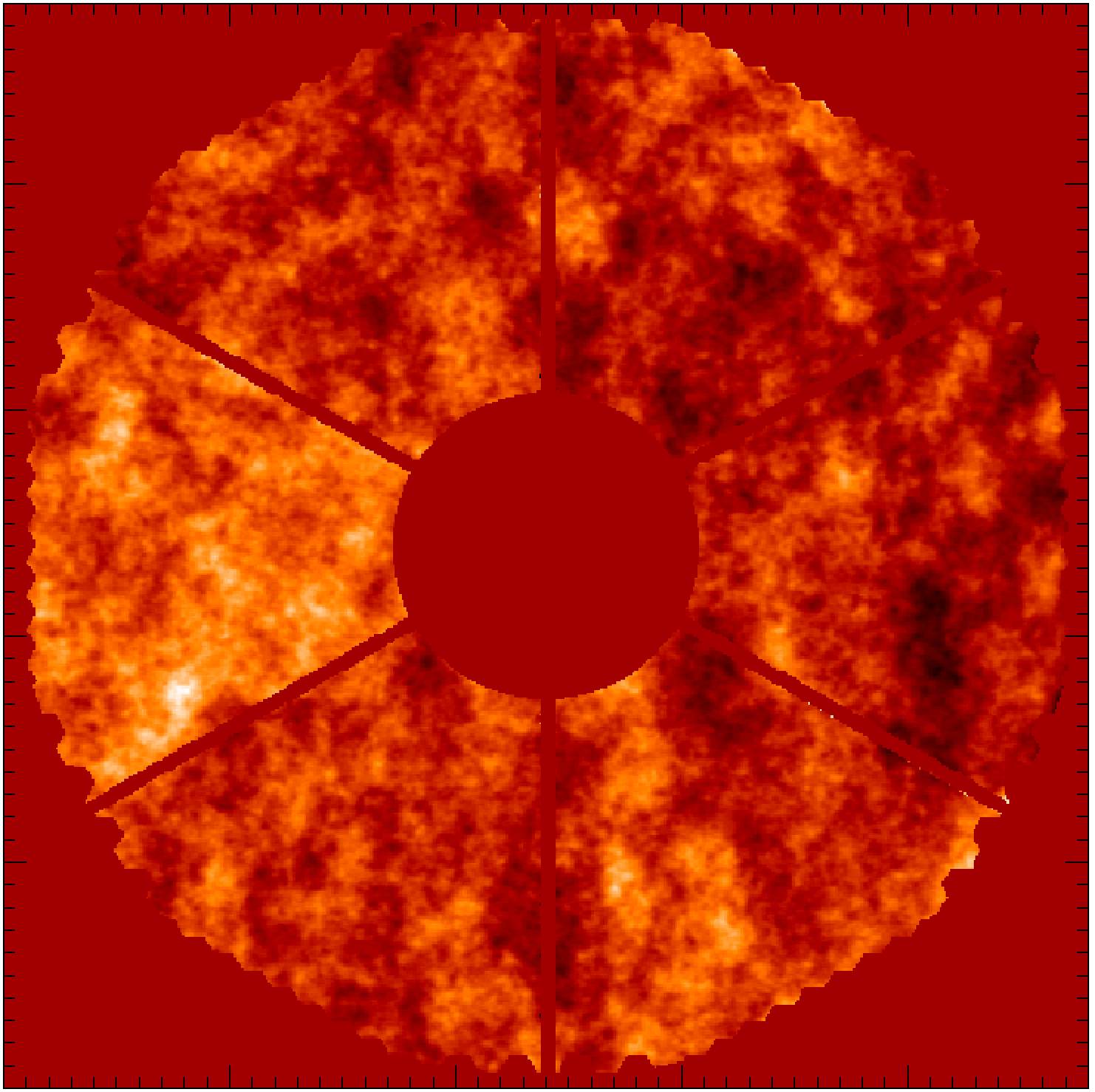}
    \caption{Residual phase from MORFEO simulation along a direction at 55arcsec off-axis in presence of residual turbulence and a petaling error on segment no. 1 of 500nm.}
    \label{fig:phase_see0.65}
\end{figure}

\begin{figure}[h!]
    \centering
    \subfigure[Seeing 0.2arcsec.\label{fig:ave_frame_see0.20}]
    {\includegraphics[width=0.42\columnwidth]{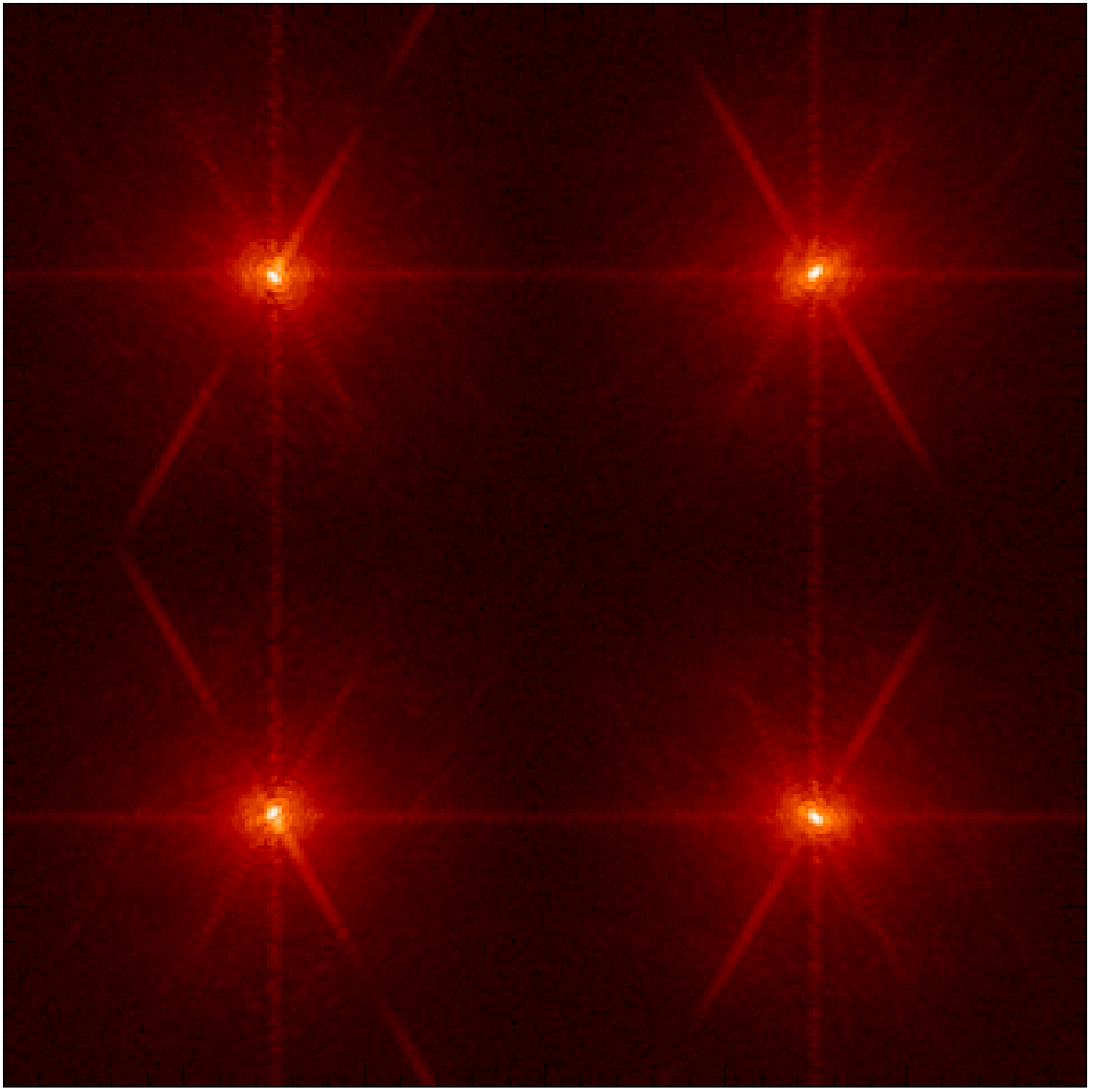}}
    \subfigure[Seeing 0.65arcsec.\label{fig:ave_frame_see0.65}]
    {\includegraphics[width=0.42\columnwidth]{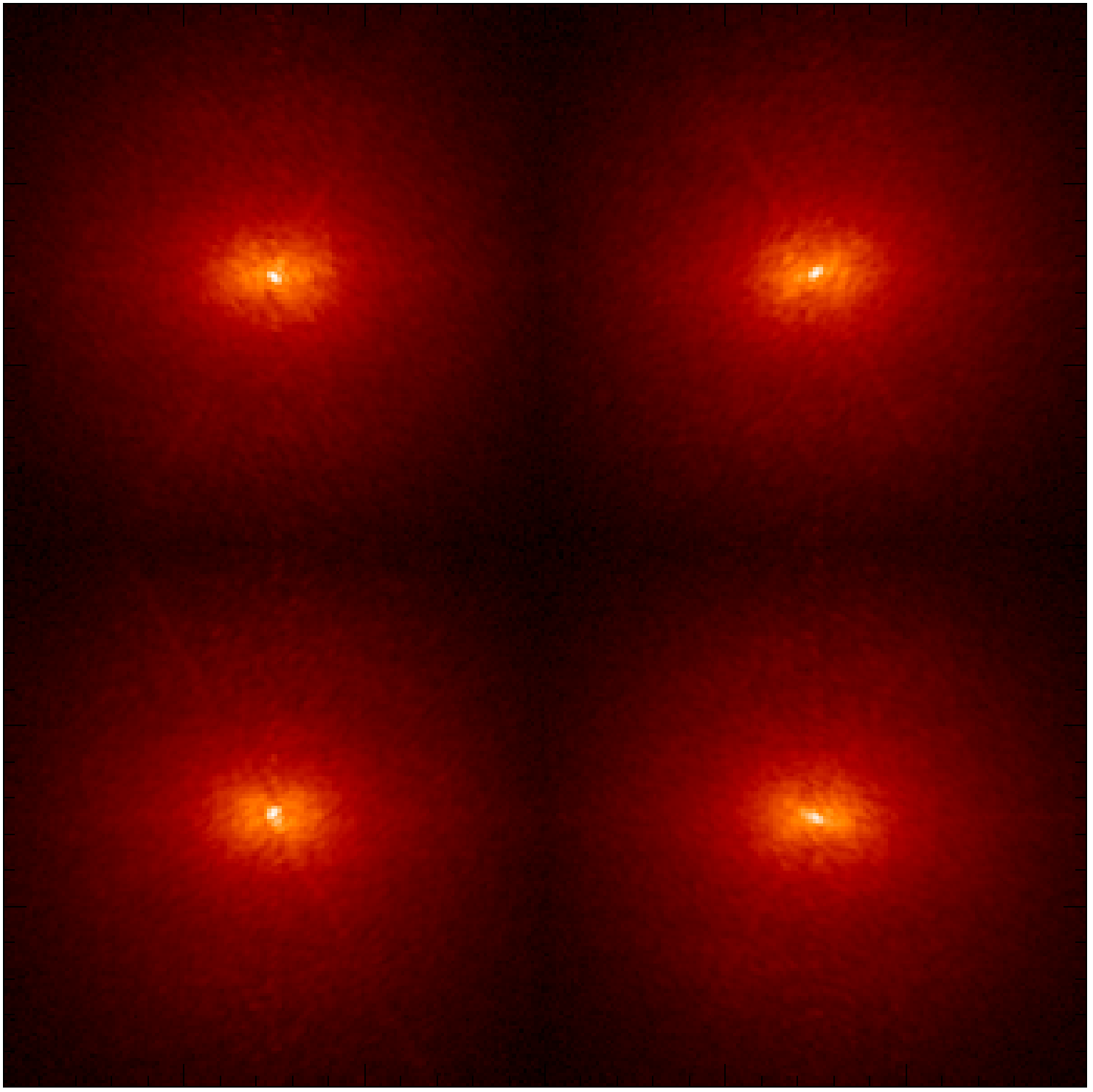}}
    \caption{Average image (2s) of the 2$\times$2 H band NGS WFS ($\sim$1.3arcsec FoV) of MORFEO at 55arcsec off-axis in presence of residual turbulence and a petaling error on segment no. 1 of 500nm.\label{fig:ave_frame}}
\end{figure}

We run Single Conjugate AO (SCAO) like simulations starting from residual phase screens of MORFEO simulations in the baseline configuration with a single post focal DM\cite{Busoni2022SPIE}.
We select an off-axis direction at 55arcsec off-axis, that corresponds to the closest on-axis direction that the NGS WFS could reach when the large MICADO FoV is used.
We run two sets of simulations, 2 seconds long, one considering a very good seeing of 0.2arcsec and one considering the median atmospheric conditions with seeing of 0.65arcsec\cite{2013aoel.confE..89S}.
In the simulation we set up a petaling error on segment no. 1 of 500nm (see Fig. \ref{fig:phase_see0.65}).
In the first case the PSFs quality is good (see Fig. \ref{fig:ave_frame_see0.20}) and the LIFT measurement is accurate with error RMS of few tens of nm, while when the seeing value is 0.65arcsec the correction has poor contrast close to the PSF peak (see Fig. \ref{fig:ave_frame_see0.65}) and LIFT is not able to measure correctly the petal error: the average estimation error on the segment where the 500nm differential piston is present is 340nm and the standard deviation is 260nm.
Unfortunately the NGS WFSs in MORFEO patrols a FoV that is far from the well corrected FoV\footnote{the correction is particularly low for the baseline configuration with a single post focal DM.} because they work in the same bandwidth of the science and they must avoid any vignetting of the science FoV.


\section{Considerations}\label{sec:consid}

LO WFS can be used as a petalometer only when AO residual is low.
Unfortunately this greatly limits the sky coverage of this sensor.
Another critical point is that the capture range will be limited $\pm \lambda/2$, with $\lambda$ = 1650nm.

\begin{figure}[h!]
    \centering
    \subfigure[Seeing 0.65arcsec and correction with ELT M4 and 1 post focal DM.\label{fig:ave_frame_see0.65_1pfdm}]
    {\includegraphics[width=0.42\columnwidth]{ave_frame_see0.65.png}}
    \subfigure[Seeing 0.65arcsec and correction with ELT M4 and 2 post focal DMs.\label{fig:ave_frame_see0.65_2pfdm}]
    {\includegraphics[width=0.4185\columnwidth]{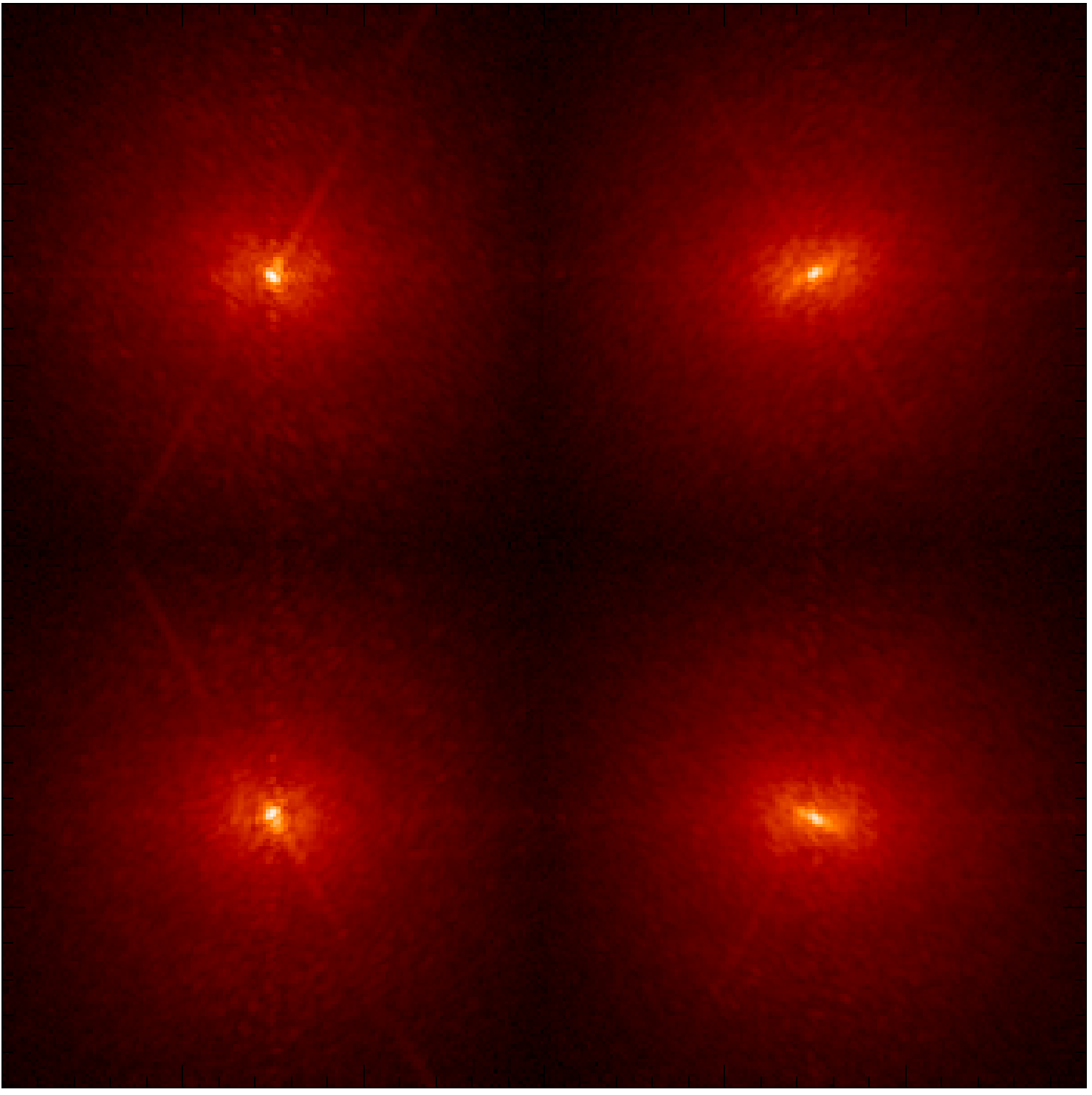}}
    \caption{Average image (2s) of the 2$\times$2 H band NGS WFS ($\sim$1.3arcsec FoV) of MORFEO at 55arcsec off-axis in presence of residual turbulence and a petaling error on segment no. 1 of 500nm.\label{fig:ave_frame_1_2}}
\end{figure}

We found that a few options are available to improve the correction:
\begin{itemize}
    \item A different algorithm for the focal plane sensing: for example the one presented in Rossi \textit{et al.}\cite{Rossi2022}.
    \item A second post focal DM. As we showed in Ref. \citeonline{2020SPIE11448E..2SA} the second post focal DM is able to improve the performance in particular in the technical FoV: we expected a ratio between 1.15 and 3 in H band SR (the first value is for good atmospheric conditions and close to the science FoV, the second value is for bad atmospheric conditions and at the outer edge of the technical FoV). An example of the average frames from NGS WFS comparing the case with 1 and 2 post focal DMS is shown in Fig. \ref{fig:ave_frame_1_2}. We replicate the simulation presented in Sec.~\ref{sec:sim} and the average estimation error on the segment where the 500nm differential piston is present is 170nm (2 times better than the 1 post focal DM case) and the standard deviation is 90nm (3 times better than the 2 post focal DM case). These preliminary numbers makes us confident that in closed loop LIFT correction will be good. Still the limited capture range issue ($\pm \lambda /2$) is not solved with this option.
    \item A dual AO\cite{1992A&A...261..677R} configuration. We have already evaluated this option for MORFEO\cite{2018SPIE10703E..11C} and it will be similar (or slightly better) of what we can gain with the second post focal DM. 
    \item A dedicated SCAO system, that means a SCAO assisted petalometer. This option is more complex, because it will require a dedicated DM and a high sampling WFS to allow the required correction. Moreover it should be able to find stars bright enough to work in most pointings. This goes in the direction to move this SCAO assisted petalometer in front of the MORFEO relay where a large FoV is available and we still benefit from M4 correction, but it is not trivial to find here the physical volume to host this sensor. Moreover we require a SCAO correction that does not spoil the characteristic of the aberration we want to measure and correct. Nevertheless this will give us the ability to measure and correct petaling errors on most of the sky in most of the atmospheric conditions.
\end{itemize}

\section{Preliminary simulations with an assisted SCAO petalometer}\label{sec:scao_ass}

Please note that we consider here as petalometer a focal plane image with LIFT, but this in principle can be any WFS able to measure petals.
We need high flux to feed SCAO and LIFT and we need a large FoV so consider to move it to the first part of MORFEO post focal relay, where it still can benefit from M4 correction (as we said in the previous section).
Moreover to get a large capture range for the petalometer we need two bandwidths (see Ref. \citeonline{Plantet2022}).
A first guess for the limiting magnitude for this sensor is H=16: there is a probability of 64\% of finding one star with such a magnitude in a radius 53-90arcsec (53arcsec as minimum to avoid vignetting MICADO) at south galactic pole.
The expected flux for a star with H magnitude equal to 16 (K5 stellar class) is:
\begin{itemize}
    \item In R+I (SCAO part) we have 0.5ph/ms on a 1$\times$1m sub-aperture with 0.3 throughput.
    \item In H (petalometer) we have 11ph/ms on the full ELT aperture with 0.3 throughput.
\end{itemize}
These fluxes are enough to have a 40$\times$40 SCAO (required to improve the WFE) and LIFT on the full aperture PSF on a couple of narrow bands (as we said before two bandwidths are required to improve the capture range\cite{Plantet2022}) working at about 100Hz.

\begin{figure}[h]
    \centering
    \includegraphics[width=0.6\linewidth]{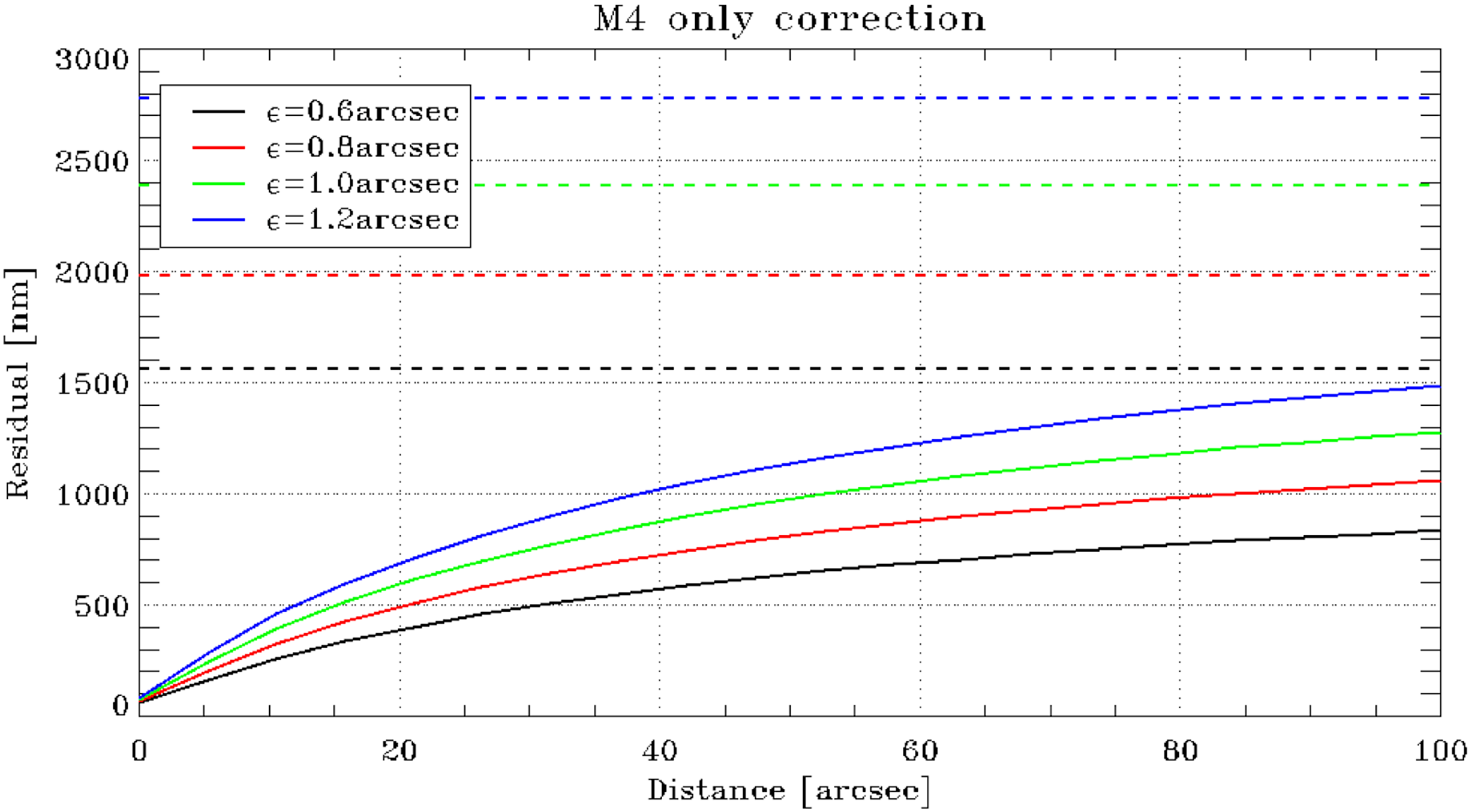}
    \caption{Expected high order residual (tilt excluded) as a function of off-axis distance when there is only M4 correction during MORFEO operation, computed with TIPTOP\cite{2021SPIE11448E..2TN} for the median atmospheric conditions\cite{2013aoel.confE..89S}. Dashed lines are error values in the seeing limited case. Gain with respect to seeing limited is similar with other atmospheric profiles.}
    \label{fig:M4_corr}
\end{figure}

We set up a simulation in this conditions considering that M4 correction gives a seeing reduction of a factor 2 or larger in high order error (see Fig. \ref{fig:M4_corr}).
So, considering a WFS with 1m sub-apertures, a framerate of 500Hz, 20ph/frame/sub-ap. and a DM with 1m pitch, we got with end-to-end simulation\cite{doi:10.1117/12.2233963}: SR(H) = [0.78,  0.61,  0.42,  0.27] for seeing [0.2, 0.4, 0.6, 0.8]arcsec respectively.
Given the expected correction from M4 and considering a initial seeing equal or smaller than 1.2arcsec, we expect to have an equivalent seeing$\leq$0.6arcsec and, consequently, a SR(H)$\geq$0.4.

\begin{figure}[h]
    \centering
    \includegraphics[width=0.6\linewidth]{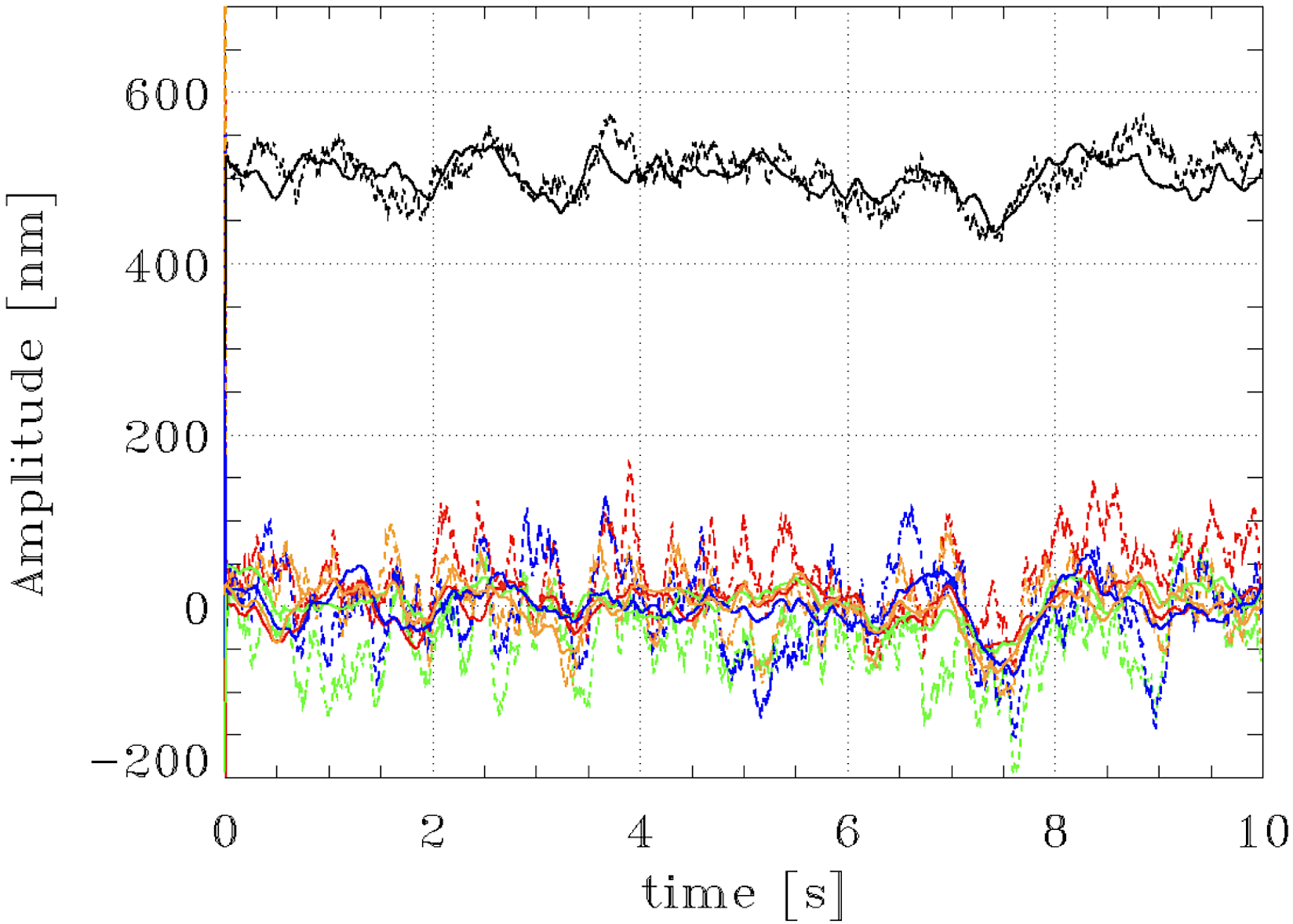}
    \caption{Time histories of the differential segment piston computed from the residual phase and SCAO assisted (LIFT based) petalometer estimation. Solid lines: average phase on segment, dashed lines: LIFT differential piston estimation.}
    \label{fig:time_hist_assisted}
\end{figure}

\begin{figure}[h!]
    \centering
    \subfigure[Focal plane ($\lambda$=1650nm, 0.5 $\times$ 0.5arcsec).\label{fig:lift_assisted}]
    {\includegraphics[width=0.419\columnwidth]{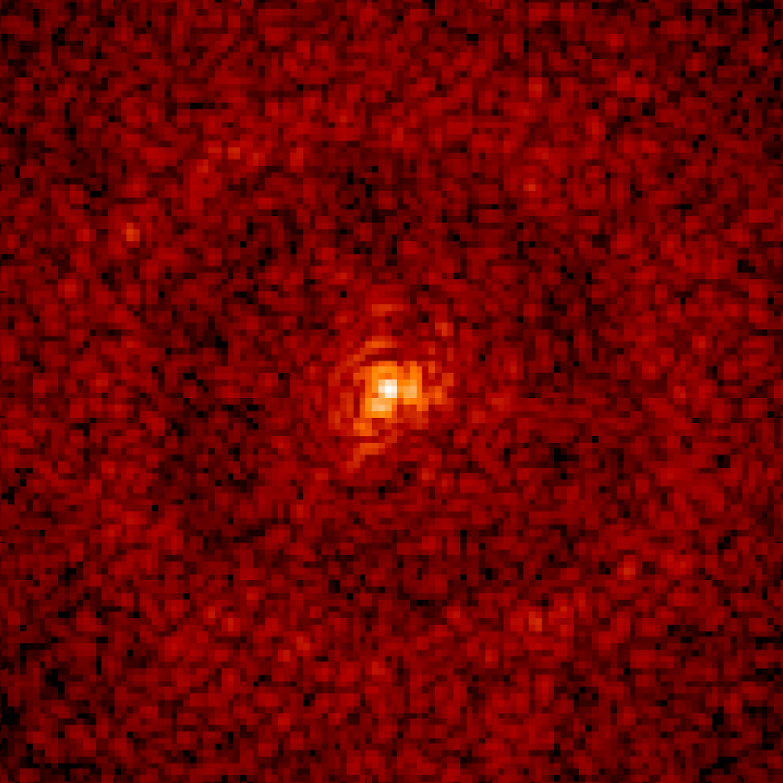}}
    \subfigure[Pupil plane (39 $\times$ 39m).\label{fig:phase_assisted}]
    {\includegraphics[width=0.42\columnwidth]{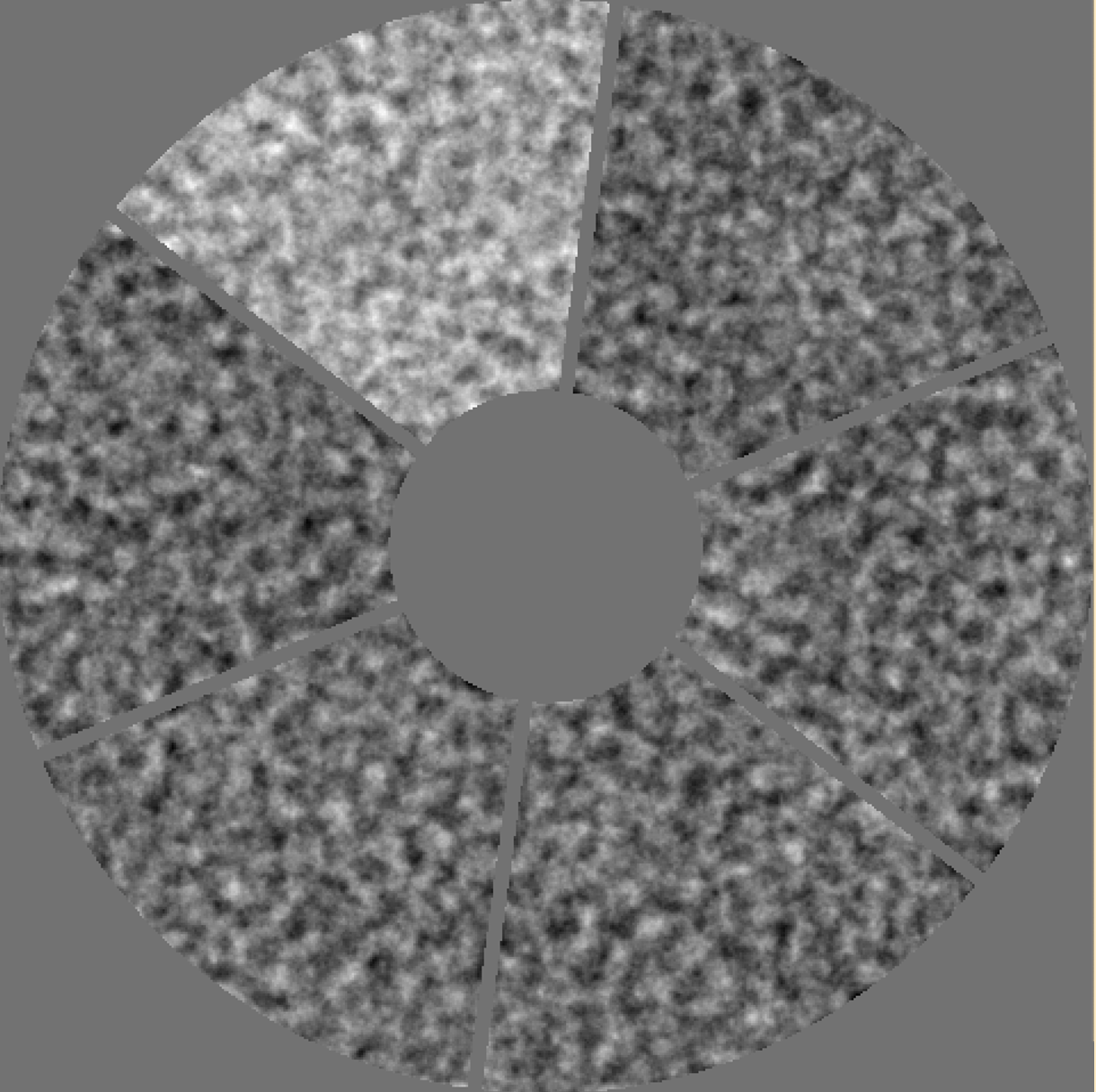}}
    \caption{One frame on the SCAO assisted petalometer with full aperture PSF.\label{fig:scao_assited}}
\end{figure}

Finally, we evaluate the capability of this phasing sensor: we run a simulation, 10 seconds long, with the SCAO system producing a H band SR of 0.4 and, as can be seen in Fig.\ref{fig:time_hist_assisted}, LIFT estimation at 500Hz in high H band flux condition (here it is done on a full aperture PSF over 40 $\times$ 40 pixels) is accurate with such correction level:  estimation error RMS at 500Hz of $\sim$50nm, st. dev. of $\sim$40nm.
As an example Fig. \ref{fig:scao_assited} shows the pupil phase and the focal plane image of one frame of the simulation.
%
%
%
%

\section{Conclusion}

We present a preliminary analysis of a possible petalometer for MORFEO showing its limitations, mainly capture range, compatibility with all atmospheric conditions and sky coverage, and a few options to improve it.
One of the most promising solution is a SCAO assisted petalometer: this is just a preliminary idea and it is a complex solution that requires extra budget and extra space.
Please note that this is the first part of the work MORFEO consortium is carrying on to find a solution for the sensing of petal phasing errors.
We do not have a comprehensive answer for the question of the title, but we can say that in some conditions the MORFEO low orders WFSs will be able to give an accurate estimation of petal errors with a capture range of $\pm \lambda/2$ (with $\lambda$ = 1650nm).
As a final remark we would like to underline the fact that the configuration of MORFEO with a second post focal DM will improve significantly the differential piston estimation with LIFT.



\bibliography{report}   
\bibliographystyle{spiejour}   

\end{document}